\documentclass[preprint,12pt]{elsarticle}

\usepackage{graphicx}
\usepackage{epstopdf}
\usepackage{multirow}

\usepackage{amssymb}

\journal{Journal of Physics G}

\begin{document}

\begin{frontmatter}



\title{Transverse-velocity scaling of femtoscopy in $\sqrt{s}=7$ TeV proton-proton
collisions}


\author{T. J. Humanic}

\address{Department of Physics, Ohio State University, Columbus, OH, USA}

\begin{abstract}
Although transverse-mass scaling of femtoscopic radii is found to 
hold to a good approximation in heavy-ion collision experiments, it is seen to fail for 
high-energy proton-proton  collisions. It is shown that if invariant radius parameters 
are plotted versus the 
transverse velocity instead, scaling with the transverse velocity is seen in $\sqrt{s}=7$ TeV
proton-proton experiments. A simple semi-classical model is shown to qualitatively reproduce this transverse velocity scaling.
\end{abstract}



\end{frontmatter}


\section{Introduction}
A common feature seen in identical-particle high-energy heavy-ion collision femtoscopy 
experiments  is transverse-mass scaling, i.e. ``$m_T$ scaling'', of the extracted radius 
parameters~\cite{Adam:2015vja,Lisa:2005dd}, where 
$m_T=\sqrt{k_T^2+m_0^2}$, $k_T$ is the average momentum of 
the particle pair, i.e. $k_T=\left | \overrightarrow{p_a}+\overrightarrow{p_b}\right | /2$, 
and $m_0$ is the particle rest mass. Transverse-mass scaling manifests itself when plotting
radius parameters extracted from identical-particle pairs of various masses from the same
colliding system versus $m_T$, and observing that the downward trend of the radius parameters
with increasing $m_T$ does not depend on the particle mass. A good example of this
for the case of one-dimensional femtoscopy where the invariant radius parameter is
extracted is seen in Fig.~8 of Ref.~\cite{Adam:2015vja}. This figure plots invariant
radius parameters versus $m_T$  for identical pion, kaon and proton pairs extracted
in Pb--Pb collisions at $\sqrt{s_{NN}}=2.76$ TeV from the ALICE experiment. As seen in this
figure, $m_T$ scaling for the invariant radius parameters holds to a good approximation
for all of the measured collision centralities. This dependence
is normally explained as a signature of radial flow in these collisions, which can be reproduced
by various models~\cite{Lisa:2005dd,Kisiel:2014upa}.

\begin{figure}
\begin{center}
\includegraphics[width=100mm]{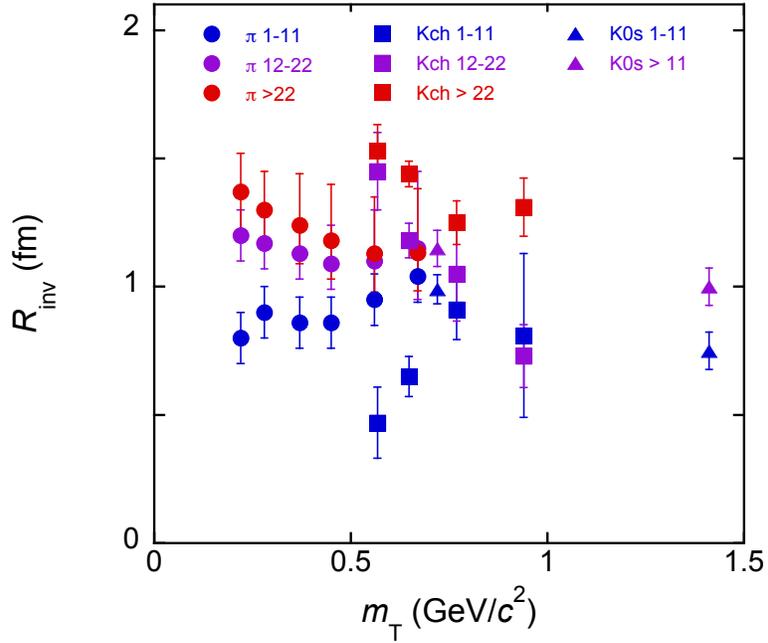} \caption{Experimental invariant radius parameters
versus transverse mass, $m_T$, for various charged-particle multiplicity ranges 
(given by the numbers in the legend) extracted from one-dimensional 
femtoscopic charged pion, charged kaon and neutral kaon 
analyses in $\sqrt{s}=7$ TeV proton-proton
collisions by the ALICE experiment~\cite{Abelev:2012ms, Abelev:2012sq}. 
This is a reproduction of Fig.~4 in Ref.~\cite{Abelev:2012sq}. (color online)}
\label{fig1}
\end{center}
\end{figure}

The situation is seen to be different in high-energy proton-proton collisions. Femtoscopic
experiments can show a similar downward trend of the extracted radius parameters with increasing
$m_T$ for a single-mass particle pair in high-multiplicity charged-particle collisions, 
but the $m_T$ scaling seen in heavy-ion experiments when
plotting the radius versus $m_T$ for several masses is not seen. Figure~\ref{fig1} shows an
example of this lack of $m_T$ scaling. The figure shows experimental invariant radius parameters
versus the average $m_T$ in that bin for various charged-particle multiplicity ($N_{ch}$) ranges extracted from one-dimensional femtoscopic charged pion, charged kaon and neutral kaon 
analyses in $\sqrt{s}=7$ TeV proton-proton
collisions by the ALICE experiment~\cite{Abelev:2012ms, Abelev:2012sq} 
(This is a reproduction of Fig.~4 in Ref.~\cite{Abelev:2012sq}). As seen in the figure,
for the higher multiplicity ranges, $N_{ch}$ $12-22$ and $N_{ch}>22$, the invariant 
radius parameter, $R_{inv}$, decreases for increasing $m_T$ individually for the pion and kaon analyses, however there is no apparent scaling between the pions and kaons with $m_T$. A feature that is common to both
pions and kaons is the overall increase in $R_{inv}$ with higher multiplicity, which is also seen in
heavy-ion collisions. This can be attributed to an increased volume of the collision producing
larger multiplicity and radii due to, for example, final-state
hadronic rescattering~\cite{Akkelin:1995gh,Truesdale:2012zz,Humanic:2013xga}. 
Another interesting feature seen in this
figure is the rising $m_T$ dependence of $R_{inv}$ for the pions and charged kaons in the lowest
multiplicity range, $N_{ch}$ $1-11$, also not showing $m_T$ scaling.

So far there has been no satisfactory explanation in the literature for this lack of $m_T$ 
scaling observed in femtoscopic invariant radius measurements in high-energy proton-proton 
collisions. 
One can at least point out the differences between high-multiplicity pp collisions and 
central heavy-ion collisions that might contribute to this difference in observed scaling.
In central heavy-ion collisions, particle production proceeds
via many soft nucleon-nucleon collisions creating thousands of particles, mostly pions,
which then undergo final-state rescattering that thermalizes the system and results
in radial flow. If the scattering cross sections of different particle species are averaged-out
in this dense medium, then all particles should participate equally in the radial flow and
the source radii are expected to approximately follow a common $m_T$ scaling for 
all particle species~\cite{Lisa:2005dd}. This situation is in contrast to particle production
in pp collisions which proceed through a single hard pp collision that produces at
most $\sim 40-50$ particles. Even for these highest-multiplicity pp collisions, significant collective
effects of the same nature as heavy-ion collisions would not be expected. Thus
it is not surprising that $m_T$ scaling is not experimentally observed in pp collisions.

The goal of the present work is to
1) show that a different scaling is seen for proton-proton collisions and 2) develop a simple 
semi-classical model to
qualitatively reproduce this new scaling.

\section{Transverse-velocity scaling}
\begin{figure}
\begin{center}
\includegraphics[width=100mm]{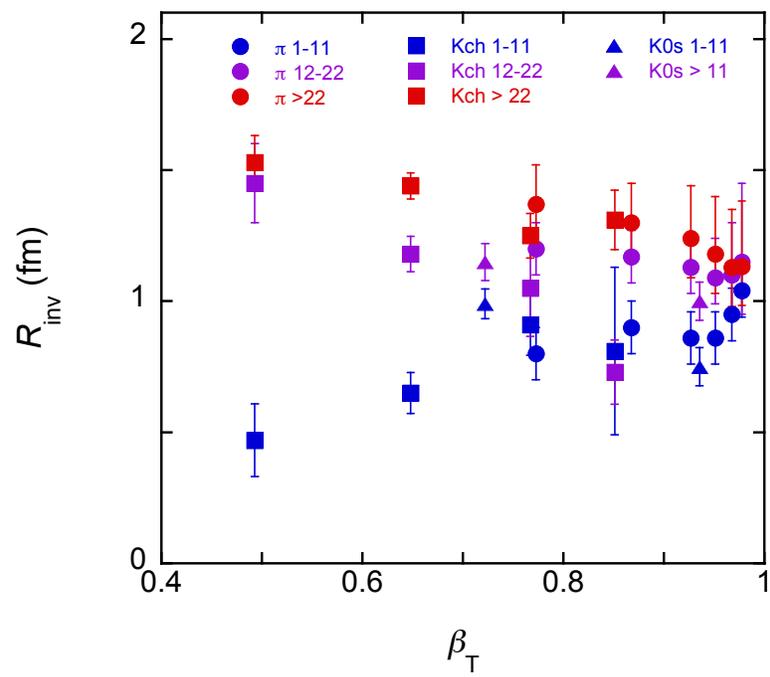} \caption{Same data as in Fig.~\ref{fig1} replotted
versus transverse velocity, $\beta_T$. (color online)}
\label{fig2}
\end{center}
\end{figure}

Figure~\ref{fig2} shows the data in Fig.~\ref{fig1} replotted versus the average transverse
velocity, $\beta_T$, of the $m_T$ bin, where $\beta_T=k_T/m_T$, and $k_T$ is the average
value in the $m_T$ bin. As seen, there now appears to be a scaling of $R_{inv}$ on $\beta_T$ for
each charged multiplicity range, i.e. the dependance of $R_{inv}$ is only on $\beta_T$ and not
on the particle mass within the uncertainties of the measurements. 

At this point, the $\beta_T$ scaling is presented as empirical, i.e. there is no theoretical 
guidance for why  this scaling should work. In the next section a simple semi-classical model is 
developed to try to describe this scaling. Results from the model are then compared with
the experimental results in Fig.~\ref{fig2}.

\section{Description of a simple semi-classical model to describe the transverse-velocity scaling }
In this section, a simple semi-classical model is described which attempts to mock up the $\beta_T$
scaling seen in the measurements in Fig.~\ref{fig2}. In addition to showing $\beta_T$ scaling
between the pions and kaons, the model also should account for the different character
of $R_{inv}$ on $\beta_T$ seen for the lowest multiplicity range, i.e. increasing
$R_{inv}$ with increasing $\beta_T$ for $N_{ch}$ $1-11$, and the higher ranges, 
i.e. larger overall $R_{inv}$ that does not vary much with $\beta_T$ for $N_{ch}$ $12-22$
and that is decreasing for increasing $\beta_T$ for $N_{ch}>22$.
It will be assumed that the higher multiplicity ranges are different from the lowest one
due to 1) longer meson hadronization times making $R_{inv}$ larger, and 2) the presence of
a significant radial flow acting before hadronization which ultimately results in the decreasing 
dependence of $R_{inv}$ on $\beta_T$ for the highest multiplicity range. 
It will be assumed that the additional 
energy stored in the radial flow is converted to additional particle production upon 
hadronization, although this is not explicitly addressed in this simple model.  
A Monte Carlo calculation is used. After hadronization, pairs of charged particles, 
i.e. charged pion or kaon pairs, are given a combined Bose-Einstein symmetrization and Coulomb
interaction weight, and a
one-dimensional correlation function in terms of the invariant momentum difference is then
formed. The radius parameter, $R_{inv}$, is then 
extracted by fitting a Gaussian spacial model to the correlation function, similar to what was done by 
the experiments that extracted the $R_{inv}$ values from data shown in Fig.~\ref{fig2}. A more
detailed description of this calculation now follows.

The space-time point of the $i^{th}$ particle of rest mass $m_{0i}$ at hadronization in the 
proton-proton collision frame 
$(x_i,y_i,z_i,t_i)$ with energy-momentum $(p_{xi},p_{yi},p_{zi},E_i)$ is determined in the model by a Gaussian distribution for the hadronization time of width $\sigma_t$, such that
\begin{eqnarray}
\frac{dn}{dt_i}\propto\exp(-\frac{t_i^2}{2\sigma_t^2})
\label{prob}
\end{eqnarray}
\begin{eqnarray}
x_i=t_i\beta_{rad,i}\cos\phi_i \ \ \ \ \ y_i=t_i\beta_{rad,i}\sin\phi_i \ \ \ \ \ z_i=t_i\frac{p_{zi}}{E_i}
\label{geom}
\end{eqnarray}
\begin{eqnarray}
\beta_{rad,i}=\frac{\beta_i+\beta_f}{1+\beta_i\beta_f} \ \ \ \ \ \beta_i=\frac{p_{Ti}}{E_i}
\label{velocity}
\end{eqnarray}
where $p_{Ti}=\sqrt{p_{xi}^2+p_{yi}^2}$, $\phi_i$ is the azimuthal angle of the $i^{th}$ particle set randomly between $0-2\pi$, and $\beta_f$ is the average radial flow velocity.
The quantity $\beta_{rad,i}$ is the relativistic sum of the radial flow velocity and
transverse particle velocity.
The quantities $\sigma_t$ and $\beta_f$ are free parameters to be adjusted
to get the best agreement with the measurements. The energy-momentum of each particle
is determined from fits of exponential distributions to ALICE experimental $p_T$ distributions
for charged pions and kaons from $\sqrt{s}=7$ TeV proton-proton collisions~\cite{Adam:2015qaa}, and from assuming a flat particle rapidity distribution in the range of $-1<y<1$, which is close to the
experimental rapidity range used. The assumption of a flat rapidity distribution in this
rapidity range is seen from experimental measurements to be a reasonably good 
approximation for both
pions and kaons for pp collisions at $\sqrt{s}=7$ TeV~\cite{Khachatryan:2010us,Khachatryan:2011tm}.

Quantum statistics and the Coulomb interaction are imposed pair-wise on charged boson pairs $a$ and $b$ by weighting them at their 
hadronization phase-space points 
$(\overrightarrow{r_a},t_a,\overrightarrow{p_a},E_a)$ and
$(\overrightarrow{r_b},t_b,\overrightarrow{p_b},E_b)$ with
\begin{eqnarray}
W_{ab}=G(\eta)\{1+\cos(\Delta\overrightarrow{r}\cdot\Delta\overrightarrow{p}-\Delta t\Delta E)\}
\end{eqnarray}
where,
\begin{eqnarray}
\Delta\overrightarrow{r}=\overrightarrow{r_a}-\overrightarrow{r_b}\ \ \ \ \
\Delta\overrightarrow{p}=\overrightarrow{p_a}-\overrightarrow{p_b}\ \ \ \ \
\Delta t=t_a-t_b\ \ \ \ \Delta E=E_a-E_b
\end{eqnarray}
and where $G(\eta)$ is the Gamow factor,
\begin{eqnarray}
G(\eta)=\frac{2\pi\eta}{\exp(2\pi\eta)-1} \ \ \ \ \ \eta=\frac{m_0\alpha}{q_{inv}}
\label{gamow}
\end{eqnarray}
and $q_{inv}=\left | \Delta\overrightarrow{p}\right | -\left | \Delta E\right |$ is the invariant momentum
difference and $\alpha$ is the fine structure constant.
The correlation function, $C(q_{inv})$, is formed by binning pairs in terms of the invariant momentum difference as the ratio of
weighted pairs, $N(q_{inv})$, to unweighted pairs, $D(q_{inv})$,
\begin{eqnarray}
C(q_{inv})=\frac{N(q_{inv})}{D(q_{inv})}
\label{cf_model}
\end{eqnarray}
Since a final-state Coulomb interaction, via the Gamow factor, is in the model between 
boson pairs after hadronization to more closely mock up the experimental conditions for the
femtoscopic analysis, a Gaussian function with the Gamow factor using the Bowler-Sinyukov
equation~\cite{Bowler:1991vx,Sinyukov:1998fc} is fitted to Eq.\ref{cf_model} to extract the
boson source parameters which are compared with experiment,
\begin{eqnarray}
C_{\rm fit}(q_{\rm inv})=a\{1-\lambda+\lambda G(\eta)[1+\exp(-q_{inv}^2R_{inv}^2)]\}
\label{cf_fit}
\end{eqnarray}
where $R_{inv}$ is the invariant radius parameter which, in principle, is 
related to the size of the boson source,
$\lambda$ is a parameter that reflects the strength of the quantum statistics effect as well as the degree
to which the Gaussian function fits to the correlation function, and $a$ is an overall normalization
parameter. Eq. \ref{cf_fit} is the same function as used by ALICE to extract $R_{inv}$
and $\lambda$ from quantum statistics in their measurements for the charged particle pairs,
except that instead of using the Gamow factor to account for the Coulomb interaction they
use a factor calculated from Coulomb waves. However, for the small radius parameters
that are extracted in these collisions, i.e. $\sim 1$ fm, the two methods of calculating
Coulomb yield almost the
same results~\cite{Bowler:1991vx}.

\section{Results and Discussion}
\begin{figure}
\begin{center}
\includegraphics[width=120mm]{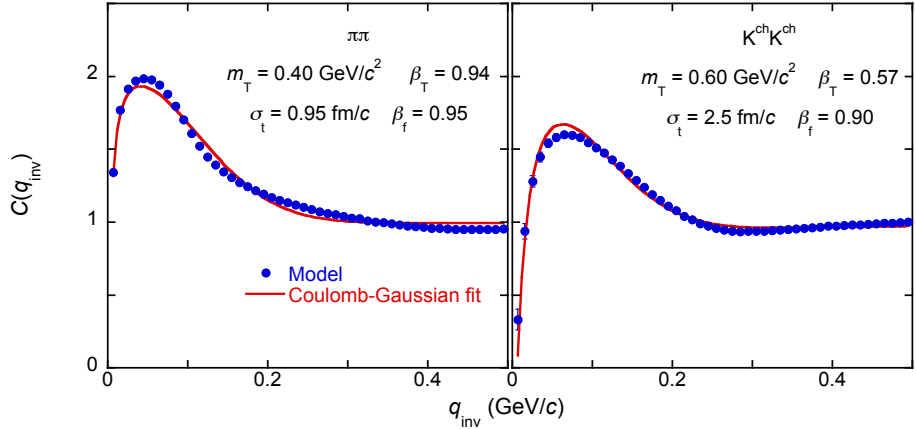} \caption{Sample correlation functions from $\sqrt{s}=7$ TeV
proton-proton collisions for charged pions and
charged kaons from the model using Eq. \ref{cf_model}. Fits of Eq. \ref{cf_fit} to the 
model points are also shown. (color online)}
\label{fig3}
\end{center}
\end{figure}

Figure~\ref{fig3} shows sample correlation functions
for charged pions and charged kaons from the model calculated from Eq. \ref{cf_model}. Fits of Eq. \ref{cf_fit} to the model points are also shown. As seen, the fits represent the model points
reasonably well.

\begin{figure}
\begin{center}
\includegraphics[width=100mm]{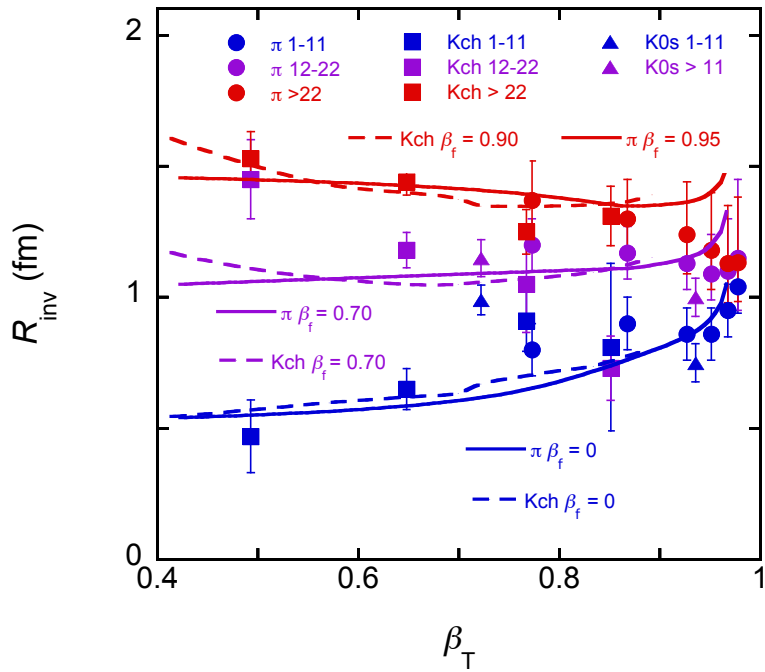} \caption{Same as Fig.~\ref{fig2} with model calculations,
shown as solid and dashed lines for pions and kaons, respectively, overlaid over the experimental points. (color online)}
\label{fig4}
\end{center}
\end{figure}

Figure~\ref{fig4} shows the same experimental results as Fig.~\ref{fig2} with the model
calculations for $N_{ch}$ $1-11$, $N_{ch}$ $12-22$ and $N_{ch}>22$ overlaid as solid lines for pions and dashed
lines for kaons. Table~\ref{tab:parameters} gives the values of the free parameters $\sigma_t$ 
and $\beta_f$ used in the calculations to adjust the model to best agree with the measurements.
The uncertainties in extracting $\sigma_t$ and $\beta_f$ are estimated to be
  $\pm 0.1$ fm/$c$ and $\pm 0.05$, respectively.
As seen, the model calculations represent the trends of the measurements reasonably well,
as well as showing the $\beta_T$ scaling between the pions and kaons. It is also seen that
$R_{inv}$ from the model sharply turns upward for $\beta_T \rightarrow 1$ for all multiplicity
ranges. This asymptotic-like behavior can be qualitatively understood by expressing $R_{inv}$
approximately in terms of the three-dimensional Gaussian radii,
\begin{eqnarray}
R_{inv}^2 \approx (R_{long}^2 + R_{side}^2 +\gamma_T^2R_{out}^2)/3
\label{3d}
\end{eqnarray}
where $R_{long}$ is the radius in the $z$-direction, $R_{out}$ is the transverse radius in the
radial direction, $R_{side}$ is the transverse radius perpendicular to the radial direction,
and $\gamma_T=1/\sqrt{1-\beta_T^2}$~\cite{Lisa:2005dd,Kisiel:2014upa}. From Eq.~\ref{3d} it is
seen that $R_{inv}$ is expected to increase asymptotically for $\beta_T \rightarrow 1$.
The experimental results for $N_{ch}$ $1-11$ also clearly show this increase for
$\beta_T \rightarrow 1$, although for the higher multiplicity ranges this behavior is not
obvious due to the larger uncertainties in the measurements there.

Looking at Table~\ref{tab:parameters}, both extracted parameters are seen to systematically
increase with increasing multiplicity for both the pions and kaons.
For the $N_{ch}$ $1-11$ model calculations, it is assumed that no radial flow is present
and thus $\beta_f$ was set to zero for both pions and kaons. The hadronization
time width, $\sigma_t$, is seen to be slightly larger for the kaons than for the pions, but
both are seen to be relatively small at 0.5 and 0.7 fm/$c$, respectively. The $\beta_f$
values for the higher multiplicity ranges are also seen to be equal within the uncertainties
for the pions and kaons, suggesting that the pions and kaons are experiencing the same 
radial-flow field for each multiplicity range.

For the $N_{ch}>22$ model calculations, very large values of $\beta_f$ for pions and kaons of 0.95 
and 0.90, respectively, are required to produce decreasing $R_{inv}$ values with increasing
$\beta_T$ to agree with the measurements within the experimental errors. The $\sigma_t$
values are larger than for $N_{ch}$ $1-11$ and $N_{ch}$ $12-22$ and are significantly 
different between the pions and 
kaons at 0.95 fm/c and 2.5 fm/c, respectively, to reproduce the $\beta_T$ scaling at the highest
multiplicity. From the model point of view, the reason for the larger $\sigma_t$ for kaons is,
due to their larger mass, the velocity of a kaon in the $z$-direction is significantly smaller
than for the average velocity of pions, resulting in the size of the source in the $z$-direction
being smaller for kaons than pions for equal values of $\sigma_t$. This can be seen
in Eq.~\ref{geom}. This has a significant effect on $R_{inv}$ since it represents a combination
of sizes in both the radial and 
the $z$-direction, as seen in Eq.~\ref{3d}.

\begin{table}
 \centering

\begin{tabular}{| c | c | c | c |}
\hline
Parameter & $N_{ch}$ range & $\pi\pi$ & $K^{ch}K^{ch}$ \\ \hline
\multirow{3}{*}{$\sigma_t$ (fm/$c$)} & $1-11$ & 0.5 & 0.7 \\
   & $12-22$ & 0.75 & 1.5 \\
   & $>22$ & 0.95 & 2.5 \\ \hline
\multirow{3}{*}{$\beta_f$} & $1-11$ & 0 & 0 \\
   & $12-22$ & 0.70 & 0.70 \\
   & $>22$ & 0.95 & 0.90 \\ 
   \hline
\end{tabular}
  \caption{$\sigma_t$ and $\beta_f$ parameters used in fitting the model to the $N_{ch}$ $1-11$,
$N_{ch}$ $12-22$ and $N_{ch}>22$ data
  in Fig.~\ref{fig4}. The uncertainties in $\sigma_t$ and $\beta_f$ are estimated to be
  $\pm 0.1$ fm/$c$ and $\pm 0.05$, respectively. }
  \label{tab:parameters}
\end{table}

This model is clearly a ``toy model'' since it is based on geometry, particle kinematics and 
several ad hoc assumptions about the proton-proton collision and hadronization of the
mesons. One could try to connect a more physical interpretation to it by assuming
that the mesons are ``quasi-particles'' or "di-quarks'' before they hadronize.  In this picture, 
when quasi-particles are initially formed from the proton-proton collision, their momentum
distributions are assumed to be similar to the momentum distributions measured in experiments. 
In a low particle multiplicity environment,
hadronization takes place in a relatively short time, i.e. as in a Lund String Model
picture used in PYTHIA~\cite{Sjostrand:2006za} and little or no radial flow will be 
present. In a high particle multiplicity environment, the quasi-particle finds itself in
a quark-gluon environment and it is assumed that a radial-flow field has been
set up. The radial flow velocity adds to the initial velocity of the quasi-particle and the
hadronization time is increased due to the extended quark-gluon environment. When
hadronization finally occurs, it is assumed that the additional 
energy stored in the radial flow is converted to additional particle production.
The hadronization time for a kaon might be expected to be longer than for a pion
due to its larger mass and requirement of a strange quark.

In spite of its limitations, the toy model seems to contain elements, i.e. simple geometry 
and semi-classical particle trajectories, that allow $\beta_T$ scaling to occur with the 
appropriate choice of free parameters. In addition, the extracted
values for $\sigma_t$ and $\beta_f$ may shed some light on the hadronization process
in these collisions.

\section{Summary}
Although transverse-mass scaling of femtoscopic radii is found to 
hold to a good approximation in heavy-ion collision experiments, it is seen to fail for 
high-energy proton-proton  collisions. It is shown that if invariant radius parameters are 
plotted versus the 
transverse velocity instead, scaling with the transverse velocity is seen in $\sqrt{s}=7$ TeV
proton-proton experiments. A simple semi-classical model is shown to qualitatively reproduce this transverse velocity scaling. The elements making up the simple model, i.e. geometry and
semi-classical particle trajectories, appear to be sufficient to allow the $\beta_T$ scaling to occur with the appropriate choices of the free parameters in the model. Clearly, it would be a good test
for a more physically-motivated and detailed model calculation to describe this scaling 
seen in proton-proton femtoscopy experiments.






The author wishes to acknowledge financial support from the U.S.
National Science Foundation under grant PHY-1614835.

\bibliographystyle{utphys}   
\bibliography{betaT_scaling.bib}

\end{document}